\documentclass[12pt]{article}
 \usepackage{longtable}
\relax
\usepackage[pdftex]{graphicx}

\usepackage{amssymb}

\newcommand{\bo}{{\bar o}}

\def\bo{{\raise.15ex\hbox{\large$\Box$}}}               

\def\face{{\raise.2ex\hbox{$\displaystyle \bigodot$}\mskip-2.2mu \llap {$\ddot
        \smile$}}}                                      

\def\leftrightarrowfill{$\mathsurround=0pt \mathord\leftarrow \mkern-6mu
        \cleaders\hbox{$\mkern-2mu \mathord- \mkern-2mu$}\hfill
        \mkern-6mu \mathord\rightarrow$}       
\def\dvec#1{\vbox{\ialign{##\crcr
        \leftrightarrowfill\crcr\noalign{\kern-1pt\nointerlineskip}
        $\hfil\displaystyle{#1}\hfil$\crcr}}}           

\def\beq{\begin{equation}}
\def\eeq{\end{equation}}

\def\beqx{\begin{displaymath}}
\def\eeqx{\end{displaymath}}

\def\beql{\begin{eqnarray}}
\def\eeql{\end{eqnarray}}

\newcommand{\bea}{\begin{eqnarray}}
\newcommand{\eea}{\end{eqnarray}}

\def\[{\left [}
\def\]{\right ]}
\def\({\left (}
\def\){\right )}

\def\+{\oplus}

\begin{document}

\hbox{\hskip 12cm NIKHEF/2008-027  \hfil}
\hbox{\hskip 12cm IFF-FM-2008/03  \hfil}

\vskip .5in

\begin{center}
{\Large \bf Non-supersymmetric Orientifolds of Gepner Models }

\vspace*{.4in}
{ B. Gato-Rivera}$^{a,b}$\footnote{Also known as B. Gato}
{and A.N. Schellekens}$^{a,b,c}$
\\
\vskip .2in

${ }^a$ {\em NIKHEF Theory Group, Kruislaan 409, \\
1098 SJ Amsterdam, The Netherlands} \\

\vskip .2in

${ }^b$ {\em Instituto de F\'\i sica Fundamental, CSIC, \\
Serrano 123, Madrid 28006, Spain} \\

\vskip .2in

${ }^c$ {\em IMAPP, Radboud Universiteit,  Nijmegen}

\end{center}

\begin{center}
\vspace*{0.3in}
{\bf Abstract}
\end{center}
Starting
from a  previously collected set of tachyon-free closed strings,
we search for \break $N\!=\!2$ minimal model  orientifold spectra which contain
the standard model and are free of tachyons and tadpoles at lowest order.
 For each class of tachyon-free closed strings  -- bulk supersymmetry,
automorphism invariants or Klein bottle projection -- we do indeed find non-supersymmetric
and tachyon free chiral brane configurations that  contain the standard  model. However,
a tadpole-cancelling hidden sector could only be found in the case of bulk supersymmetry.
Although about half of the  examples we have found make use of branes that break the bulk  space-time 
supersymmetry,  the resulting massless open string spectra are nevertheless supersymmetric in all cases.
Dropping the requirement that the standard model be contained in the spectrum, we find chiral tachyon and
tadpole-free solutions in all three cases, although in the case of bulk supersymmetry all massless spectra are
supersymmetric.  In the other two cases we find truly non-supersymmetric spectra, but a large fraction of
them are  nevertheless partly or fully supersymmetric at the massless level.

\vskip 1in

\noindent
October 2008

\newpage


In a previous paper \cite{GatoRivera:2007yi}
we have obtained a large set of non-supersymmetric but tachyon-free
closed string theories from tensor products of $N=2$ minimal models.  We begin with a brief
summary of the main results of that paper.

As a first step we considered all 168 combinations  (``Gepner models" \cite{Gepner:1987qi}) 
of these minimal models, and
all possible extensions of their chiral algebra, and we 
checked the presence of tachyonic states in the resulting representations, which may serve as
chiral halves of closed string theories.

More precisely, we considered tensor products of a covariant NSR model including superghosts and
a number of $N=2$ minimal models with total central charge 9. There are 168 ways of obtaining this total.
In order to impose world-sheet supersymmetry the chiral algebra of each tensor product is extended
with alignment currents, which are spin-3 currents build out of all possible pairs of the world-sheet
supercurrents in each of the factors. The resulting CFT may then be extended by any other set of integer spin
simple currents. Typically, there are of the order of ten to a few hundred possibilities for these currents, for each tensor product, including at least one that has spin 1 and is a space-time spinor. This is the current that
imposes a GSO-like projection, which in its turn implies space-time supersymmetry. This current is unique up
to charge conjugation in each factor.

The  set of characters one obtains in this manner can be used in two ways as building blocks for
string theories. One may either use it as a fermionic sector of a closed string theory, or one may
replace the NSR model by a bosonic CFT with identical modular properties, and use the characters to
build a bosonic sector of a closed string theory. 
Combining  these sectors in all possible ways gives
rise to bosonic, heterotic or type-II closed strings. 

It is convenient to use the bosonic string language for the description of the characters. This 
``bosonic string map" was
first exploited in \cite{LLS} in a construction of chiral four-dimensional heterotic strings from self-dual
lattices ("the covariant lattice construction"), and later also by Gepner  \cite{Gepner:1987qi} 
in his famous construction of
heterotic models. In this description, the NSR factor is represented by a $D_5 \times E_8$ level-1 affine Lie
algebra. In the fermionic interpretation, only the vector and one of the spinor characters of $D_5$ are
important. The former gives rise to space-time scalar ground states, and the latter to space-time
spinor ground states. In the left-moving sector of the closed string, their masses
are correctly given by the bosonic string mass formula, and are equal to $h-1$, where $h$ is the
conformal weight of the ground state. In the bosonic interpretation all $D_5$ characters are relevant,
and all give space-time scalar ground states.

In the bosonic interpretation, a character is tachyonic if the conformal weight of its ground state satisfies  $h < 1$. In the
fermionic interpretation, it is tachyonic if $h < 1$ and if it is a $D_5$ vector. World-sheet supersymmetry
prohibits $D_5$ spinors to be tachyonic, and $D_5$ scalars do not correspond to physical states. If we
impose world-sheet supersymmetry when using the bosonic interpretation (which is not necessary, but
true by construction for the characters discussed in \cite{GatoRivera:2007yi}),  it
follows that also in that case spinors of $D_5$ cannot give rise to tachyons. 

In the bosonic interpretation of the NSR characters there is always at least one tachyonic character,
the vacuum with $h=0$. In the fermionic interpretation any tachyonic state must
be a $D_5$ vector and hence the minimal value for $h$ is $\frac12$. This minimal tachyonic state
is projected out automatically by any extension of the chiral algebra that is a $D_5$ spinor. In the
bosonic interpretation, one can have any number of additional tachyonic characters with $0 < h < 1$; in the
fermionic interpretation, there can be any number with $\frac12 < h < 1$. 
The first conclusion obtained in \cite{GatoRivera:2007yi}  is that for any chiral algebra extension there is at least one
$D_5$ vector tachyon that survives all projections imposed by the chiral algebra, except if the 
chiral algebra contains the space-time supersymmetry current. This is an empirical result valid
for superconformal CFT's built out of $N=2$ minimal models. We are not aware
of any theorem that proves this result for any superconformal algebra, but it seem reasonable to conjecture that
this might be true in general.
 If such a theorem could be proved,
it would provide a precise version of the often-heard misconception that ``absence of tachyons requires
supersymmetry".
However, this would still be a misconception since it only refers to a chiral half of a closed string theory.

There is a variety of ways to use these characters to build non-supersymmetric tachyon-free string theories.
In order to avoid tachyons, the most obvious closed string construction, namely a diagonal 
pairing of
the two tachyonic chiral halves into orientable closed strings, must be avoided. However, one
may consider non-trivial MIPFs (Modular Invariant Partition Functions) and/or consider unoriented strings.
This  leaves us with the following possibilities.
\begin{enumerate}
\item{Type-II strings with off-diagonal pairings of fermionic sectors. Note that the MIPF must include a non-trivial
automorphism of the fusion rules. Otherwise it would be equivalent to a pure extension, which empirically does  not work, 
as explained above.}
\item{Heterotic strings. Here the idea would be to take one of the aforementioned type-II strings and map
one of its fermionic sectors to a bosonic one.}
\item{Type-I strings with a Klein bottle projection that removes the tachyons. The addition of the Klein bottle
introduces crosscap tadpoles, and an open string sector must be introduced to cancel these, without
introducing open string tachyons. The open string tachyons may either be avoided altogether,
or removed by the Moebius strip projection. The latter option is however available only for rank-2
tensor matter, not for bi-fundamentals.}
\item{Type-I strings with a space-time supersymmetric MIPF removing the closed string tachyons. 
Within the closed sector this is a trivial solution, but  supersymmetry might be violated in 
an orientifold theory by the open string sector \cite{Antoniadis:1999xk}. Since supersymmetry is only introduced as a MIPF
extension, the set of characters in which the open sector is expanded is non-supersymmetric and
in general tachyonic. Open string tachyons must be avoided as in the previous case.}
\end{enumerate}

 In \cite{GatoRivera:2007yi} options 1, 3 and 4 were explored, and many examples were found.
  However the
 open sector, which is optional in case 1, obligatory in case 3 in order to cancel tadpoles, and needed in case 4 to get a non-supersymmetric result, was not considered, and will be the subject of the present paper. The scope of the search done in  \cite{GatoRivera:2007yi} was limited for practical reasons, and we will use the same limitations here, plus a few additional ones. 
 We do not consider the $(2,2,2,2,2,2)$ tensor product, because even though it has
a huge number of tachyon-free MIPFs \cite{GatoRivera:2007yi}, it is known to have an extremely low success rate in the supersymmetric case \cite{Dijkstra:2004cc, adks,kst}. 
 Furthermore we limit the number of boundary states to 1750.
  This limit is needed in order  to allow us to search for standard model configurations, which
 grows with the fourth power of the number of boundary states. This  search will  be done using the general method proposed in \cite{adks},
 where it was applied to the supersymmetric case with the same limit of 1750 boundary states.
 This limit comes on top of the limit to 4000 primaries used in \cite{GatoRivera:2007yi}.
  We will not consider
 zero tension orientifold planes, because they cannot provide a solution to the dilaton tadpole condition, and we also omitted MIPFs with less
 than 5 boundary  states (counting complex ones as two), because they can never produce the standard model.
 Finally, in the case of the $1^9$ tensor
 product we did take into account all permutation symmetries, which reduces the number of distinct possibilities with respect to   \cite{GatoRivera:2007yi}, where only a subset of the permutations was taken into account.
 In total, we have considered 10635 MIPFs of type 1 (with a total of 66336 orientifolds), 2998 of type  3 (with a total of 9075 orientifolds
 with the required Klein bottle)
 and 15372  MIPFs of type 4, with 95008 orientifolds.

 In theories of type 1, 3 and 4, the open sector offers
 the only way to get a string spectrum that resembles the observed particle spectrum of the
 standard model. Indeed, the best possible outcome is a rather attractive one: exactly the standard model
 spectrum. Our goal is to find out how close we can get to that spectrum within the context of rational conformal
 field theory, {\it i.e.} with exact perturbative string theory. 
 
 Before addressing that question, let us consider option 2, which offers in principle the same possibility. In general,
 heterotic strings provide a natural way to get rid of tachyons in string theory, precisely because by construction
 they are non-diagonal. Indeed, many examples have already been found for heterotic strings, see {\it e.g.}
 \cite{Dixon:1986iz,AlvarezGaume:1986jb,Kawai:1986ah,LLS,Dienes:2007ms}.
   However, the kind of heterotic strings one gets from Gepner models are not the most
 promising ones. The  bosonic string map relates any such heterotic string to a type-II string. Any tachyon
 of that type-II string automatically appears as a tachyon (in the vector representation of $SO(10)$) in
 the heterotic string. But in addition to that, the heterotic string also has tachyons from tachyonic singlets
 in the bosonic chiral half, which are not physical states if this chiral half is interpreted fermionically. Hence
 only a subset of the bosonically mapped theories of the first type will be tachyon-free.
 
 We have examined the heterotic interpretation of the 10635 orientable tachyon-free type-II MIPFs and found that 4513 of them
 are also tachyon-free as heterotic strings. This is a typical example of such a spectrum:
\begin{eqnarray*}
\hbox{Left-handed fermions:}\  \  &304 \times (1) + 32 \times (16) + 40 \times (10) + 8 \times (16^*) \\
\hbox{Scalars:}\ \ \ \ &778 \times (1) + 40 \times (16) + 108 \times (10) + 40 \times (16^*) 
\end{eqnarray*}
This example has a net number of 24 chiral families in the 16 of $SO(10)$. In the other examples, the net number of families
is often zero, and usually a multiple of 6 and/or 4, a well-known feature of the supersymmetric Gepner models. We did not encounter
any cases with 3 families, and the minimal number of families we found was 6. 

The  class of heterotic partition functions considered here is based on symmetric MIPFs. 
Only symmetric MIPFs were considered
in  \cite{GatoRivera:2007yi}  because they were collected as a first step towards orientifold model building.   With asymmetric MIPFs
the possibilities for getting tachyon free heterotic strings are probably better. Indeed, almost two decades ago such MIPFs were already
considered in  \cite{Schellekens:1989wx}. Although only supersymmetric theories were built, this does include cases where
the bosonic sector of the heterotic string does not have the equivalent of a GSO projection. Interchanging the r\^ole of the left and
right sector  then yields a non-supersymmetric heterotic string theory with an $E_6$ gauge group. The $E_6$ is a remnant of the GSO
projection \cite{LLS,Lerche:1988zz}, but its presence  in the bosonic sector does not guarantee absence of tachyons as  it does in the fermionic sector. This is  because $E_6$ singlets can now give  rise to tachyons, whereas they are unphysical in the fermionic
interpretation. Nevertheless,  it is reasonable  to expect that some of these theories will be tachyon-free, but we will not pursue that
question further here. Although we can get thousands of non-supersymmetric
tachyon-free spectra from heterotic Gepner models,
it seems clear that genuine heterotic models with unrelated left and right sectors should have a much better success rate. 
Unfortunately that is hard to do with $N=2$ minimal model building blocks because of the requirement of modular invariance,
but  with free fields this is much easier, as was demonstrated in \cite{Dienes:2007ms}.

This concludes our remarks on the heterotic  case, and we turn now to the main subject of this paper, orientifolds. 
First we discuss the most ambitious goal, namely finding standard model spectra.  We follow the same strategy as in  
\cite{adks}, namely to find first a set of at most four boundary states producing the required spectrum, and then to  find
a hidden sector to cancel the remaining tadpoles, if needed. In both steps we impose the additional requirement
that there are no open string tachyons, not in the observable sector, nor in the hidden sector, and also  not  in the
matter that is charged under both sectors. Furthermore, we require cancellation of {\it all} tadpoles, not just the
RR-tadpoles that should be forbidden for reasons of consistency, as was done in \cite{Ibanez:2001nd}.

One could take the point of view that this is a bit too restrictive. Tachyons and tadpoles that only affect the stability of a configuration
might be ignored at this stage. Their presence might only indicate that one has landed in an unstable point in the potential. Furthermore
tadpoles are only  avoided in lowest  order of perturbation theory. At higher orders, they are essentially certain to reappear, requiring
further adjustments of the solution. Indeed, if one takes  this point of view, the non-supersymmetric Gepner models become
a huge laboratory for studying open string statistics. If tachyons and NS tadpoles are ignored, the  problem of  finding the  standard
model becomes as easy as in the supersymmetric case, but with a number of MIPFs, and a number of boundary states per MIPF,
that is  one or two orders of magnitude larger. For  standard model realizations with four boundary  states, the most common ones,
that enlarges the total number of possibilities by  five to ten orders of magnitude. If it can be  made plausible -- for example by
studying subsets -- that  the presence of tachyons or tadpole instabilities does not affect distributions of quantities of interests (for
example chiral features of possible models or the number of families), then we can drastically enhance the statistics with respect to
the supersymmetric case, and perhaps find  some rare examples that did not appear in that case.

However, this is not the  point  of view we will adopt here. Our goal is to see how close one can get to the observed standard
model spectrum within the context of exact tachyon and tadpole-free RCFT.  The definition we will adopt for the standard
model is the same, very broad one used in \cite{adks}. We require a Chan-Paton gauge group built out of at most four
(real or complex) boundary states, that contains $SU(3)\times SU(2) \times U(1)$ with a massless $Y$ boson, and
 with a SM-chiral spectrum
consisting only of three standard model families.  Here SM-chiral means ``chiral with respect to $SU(3)\times SU(2) \times U(1)$". 
The definition allows  matter that  is  chiral with respect to  some extension of the standard model, but  becomes non-chiral
when the Chan-Paton group is reduced to the SM group, as well as matter that is entirely non-chiral with respect to the full
Chan-Paton group. 

To classify the chirally distinct solutions we can make use of the same criteria and the same database used in  \cite{adks}. 
This work yielded a list of 19345 spectra that had different Chan-Paton groups, or matter that is chirally different with
respect to the Chan-Paton group, or a different massless $U(1)$ vector boson in addition to  $Y$. All of these are independent 
of the superpartners of the (M)SSM, and are therefore equally usable for the SM.  Each non-supersymmetric tachyon-free
spectrum that we have encountered in the present search is  assigned an identification number  referring to  the list of 19345 spectra of
\cite{adks}, or a new  number if it
was not seen before. Only 302 new spectra have been found in comparison to the supersymmetric case. 

The total number of  standard model spectra we have found (prior to attempting to find a tadpole cancelling  hidden sector) is
3562068. This may  be compared to the total of about 145 million found in \cite{adks}. The total number of MIPFs considered 
in the latter paper was about 4500, whereas in the present paper we have examined about 30000 MIPFs.  The success rate
per MIPF is thus about 30.000 in the supersymmetric case, and just slightly more than 100 in the non-supersymmetric case.
Note that all the supersymmetric models found in \cite{adks} would eventually also emerge in the present case, if we were to increase
the maximal number of boundary states. Even though we exclude supersymmetric extensions explicitly, there are MIPFs corresponding
to the same extensions that we {\it do } allow, in order to be able to find examples with supersymmetry in the bulk, but perhaps
not on the boundary. However,  if all the boundary states in a given model  respect space-time  supersymmetry we get a spectrum
that  would also be realizable by means  of a supersymmetric extension. In practice one would not expect to find many of the
145 million supersymmetric models of \cite{adks}, because most of the supersymmetric MIPFs are extensions of  non-supersymmetric
ones with a huge number  of boundary states.

The aforementioned tachyon-free spectra are divided in the following way over the  three  different possibilities: the vast majority, 3495302, or about 98.1\%.  occurred
for closed strings with a MIPF with a bulk supersymmetry extension; 66378, or about 1.8\% occurred for orientable automorphism MIPFs,
and only 388 cases were found for closed strings with a Klein bottle projection removing the tachyons.

The next step is to try and find a hidden sector that cancels all tadpoles and does not introduce any
tachyons. This was indeed possible, and we found a total of  896 solutions. Here we did allow more than one solution
per model  type, unlike in the search in  \cite{adks}, where no further attempts were made  if a solution had already been found
for one of the 19345 types. It turns out that  all 896 occur for the case of bulk supersymmetry, {\it i.e.} a success rate of about .03\%.
This may  be compared with the supersymmetric results of \cite{Dijkstra:2004cc}, where the average success rate is about  3\%.
If the success rate  were the same for
all three possibilities, one  would have expected 16 solutions for orientable tachyon-free automorphisms, and none for tachyon-free
Klein bottles.

The fact that all solutions occurred for the case of a supersymmetric bulk extension raises the possibility that perhaps
supersymmetry is also preserved on the boundary. We have verified that in any case all massless spectra are supersymmetric, {\it i.e} 
all bosonic representations occur with equal multiplicities as the fermionic ones, if we subtract from the latter the would-be
gauginos. The fact that the massless spectrum is exactly  supersymmetric  is not sufficient to prove that the theory is indeed supersymmetric,
but it is sufficient to conclude  that we did not achieve our goal of finding a non-supersymmetric, tachyon-free standard  model spectrum.

A further step towards answering the question whether these solutions are all supersymmetric is to examine  if the boundaries
preserve supersymmetry.  In 
452 of the 896 cases that is indeed true. This means that those 452 spectra can be realized entirely in terms of a supersymmetric
extension of the chiral algebra, and hence they belong to the class already studied in \cite{Dijkstra:2004cc,adks}. The remaining cases  
are different however.  Here {\it all} boundaries used in building the standard model and the hidden sector break space-time
supersymmetry. To be precise, boundaries are labelled by a set $[i,\psi]$ where $i$  is a representative of a  simple current
orbit, and $\psi$ a degeneracy label  \cite{Fuchs:2000cm}.  If we denote the spinor current that imposes space-time supersymmetry  by $S$,
then the monodromy of $S$ with respect to $i$ is $\frac12$. This means that if we extend the chiral algebra by $S$ (as  opposed to
just having it in the chiral algebra of the MIPF), then $i$  is projected out. This implies in any case that these examples will
not be found in a search for purely supersymmetric models starting from a supersymmetric extension of the chiral algebra.

But  are these examples supersymmetric? By  inspection of a few cases, we conclude that the supersymmetry of the massless
spectrum appears to extend to the full spectrum. In other words, although boundary states that do not preserve supersymmetry are used,
the full open string spectrum is nevertheless expressible  in terms of supersymmetric characters. Presumably supersymmetry is
realized on the boundaries with a non-trivial automorphism, as explained in \cite{SBB,RS}. It is an open question at this point if also
the interactions of these theories are fully supersymmetric, and whether they can be reformulated in terms of the explicitly
supersymmetric theories already studied in \cite{Dijkstra:2004cc,adks}. Although these examples are formally  outside the
scope  of  \cite{adks}, their spectra look  quite similar, and in particular they have the same standard model  configurations
(out of the list of 19345 spectra) that already occurred for supersymmetric MIPFs of the same tensor product. This, in combination
with the non-trivial automorphism type  of the supersymmetry realization, suggests that they may be T-duals of already known
supersymmetric models, analogous to the examples discussed in \cite{Huiszoon:2000ge}. If that is indeed the case, these
spectra would {\it not} provide examples of brane supersymmetry breaking   as discussed in 
\cite{Antoniadis:1999xk}.
This issue can be studied more explicitly,
but this is beyond the scope of the present paper.

It should be noted that the non-supersymmetric models presented  in \cite{Ibanez:2001nd} are based on a bulk theory
with $N=8$ supersymmetry, {\it i.e.}  a torus. Hence they are of the type discussed in the foregoing paragraphs, except that
only Ramond-Ramond tadpoles were cancelled in \cite{Ibanez:2001nd}. Presumably that is why the construction
of those examples was possible. 

In any case, fully supersymmetric spectra were not what we have been looking for.
It is now natural to ask if there exist {\it any} tachyon and tadpole-free non-supersymmetric models at all in this context. To investigate
that we have examined  the tadpole equations without imposing the condition that the spectrum should include the standard model.
We have considered for each orientifold choice all combinations of at most four boundary states, and collected at most one solution
per orientifold. For the case of non-supersymmetric tachyon-free automorphism MIPFs, we have found a solution for 18938 out of
the 66336 orientifolds. For the non-supersymmetric tachyon-free Klein bottles these numbers were 795 out of 7095. Finally, for
MIPFs with bulk supersymmetry these numbers were 72719 out of 95008.  

The open string spectra obtained for non-supersymmetric bulk theories may be accidentally supersymmetric,
but clearly the complete theory is not. But, as discussed above, if there is bulk supersymmetry one has to worry
if supersymmetry
is really broken. This does {\it not} seem to be the case: all 72719 spectra have equal 
numbers of fermions and bosons, after subtracting gauginos from the fermions. As in the standard model search,
we find cases where all boundaries are explicitly supersymmetric, 
and cases where the boundary states  break supersymmetry (or realize it via a  non-trivial automorphism).
There is one novel feature: in addition to monodromy charge 0 and $\frac12$, we now also find boundaries that
have monodromy charge $\frac14$ and $\frac34$  with respect to $S$.  The 
absence of any examples with bulk supersymmetry but with explicitly non-supersymmetric open string spectra 
 may be
due to  statistical reasons, but since we did find non-supersymmetric spectra in the smaller samples of tachyon-free automorphisms and
Klein bottle projections, this 
suggest that perhaps brane supersymmetry breaking  (as discussed in \cite{Antoniadis:1999xk}) cannot be realized within the context
of rational CFT and cancellation of all tadpoles.
However, this is not generally true for any kind of bulk symmetry. For example, in \cite{adks} open string spectra were
found that  are chiral and have $N=1$ supersymmetry, 
even though the bulk type-II  theory had extended supersymmetry ($N=4$ or $N=8$). 

In the other two cases there was  a surprisingly large number  of accidentally supersymmetric massless spectra.
In the automorphism case, 4818 of the 18938 solutions were  accidentally supersymmetric; in the Klein
bottle case there were 228 out of 795. We have checked how these results were influenced by the
requirement that the open string spectrum be free of tachyons. After removing that requirement, 
the total number of solutions increased from 18398 to 21290, and the number with accidental supersymmetry from 4818 to  5137
in the automorphism case. This implies that there are examples with accidental supersymmetry
for the massless sector  {\it and} tachyons! The overall
change is however quite small, and hence we have to conclude that the large amount of accidentally supersymmetric cases
is not explained by the requirement of absence of tachyons. For the tachyon-free Klein bottle  case the changes were
even smaller:  an increase from 795 to 815 solutions, out of which 229 were accidentally supersymmetric.

Even non-supersymmetric solutions are often almost supersymmetric. In figure 1 we show the spectra distributed
according to the percentage of supersymmetric multiplets (as before, after subtracting gauginos from the fermions). 
The n$^{\rm th}$ bin shows the number of solutions with at least $10 \times n\%$ supersymmetry and less than $10 \times (n+1) \%$.
The last bin only contains the cases with $100\%$ supersymmetry.  Figure 2 shows the same for tachyon-free Klein bottles.

\begin{figure}
\begin{center}
\includegraphics[width=13cm]{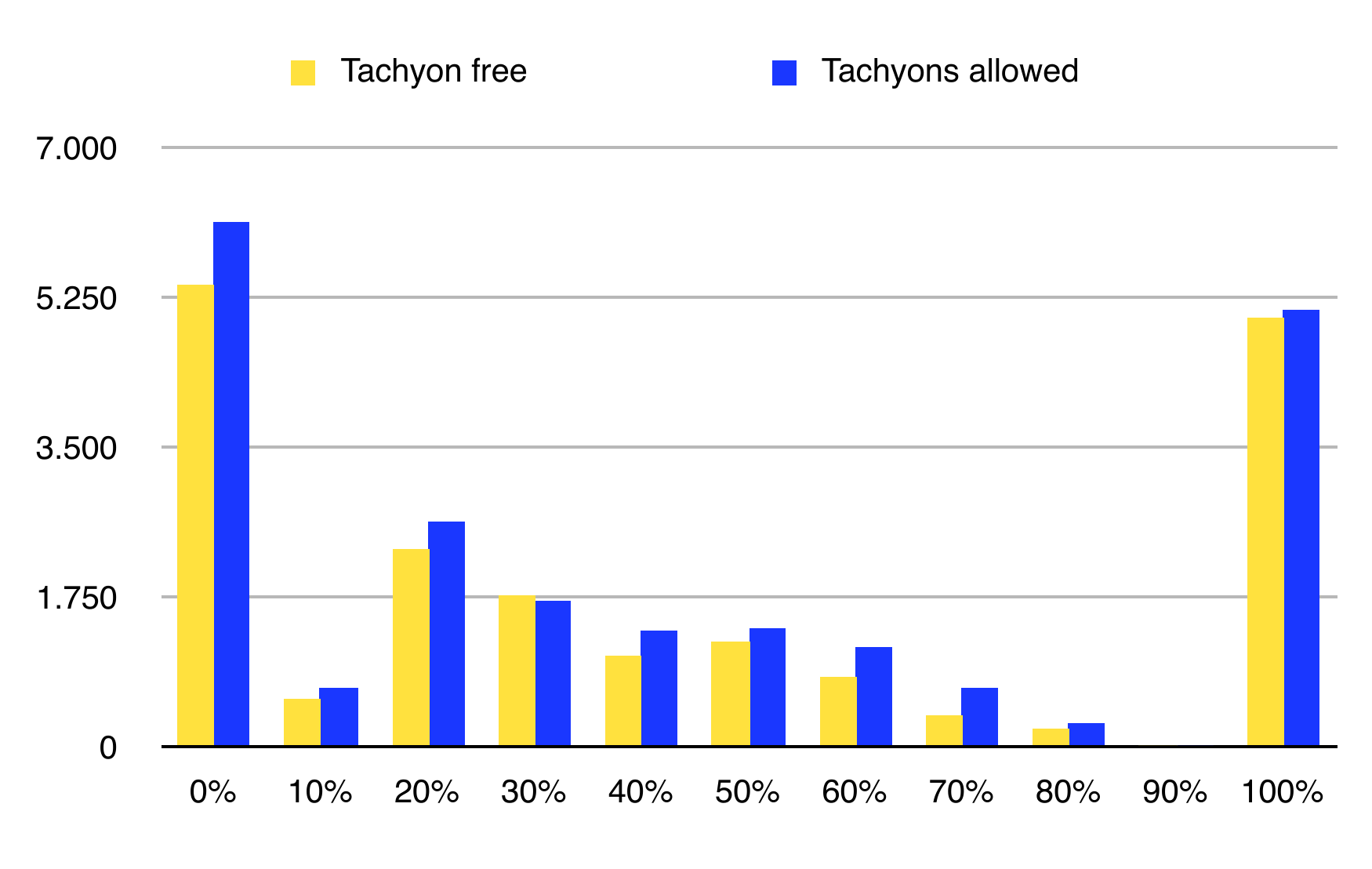}
\label{fig1}
\caption{\em Number of tadpole-free spectra with a given percentage of supersymmetric multiplets. The last bin
consists exclusively of supersymmetric spectra. This plot is for automorphism bulk invariants.}
\end{center}
\end{figure}

\begin{figure}
\begin{center}
\includegraphics[width=13cm]{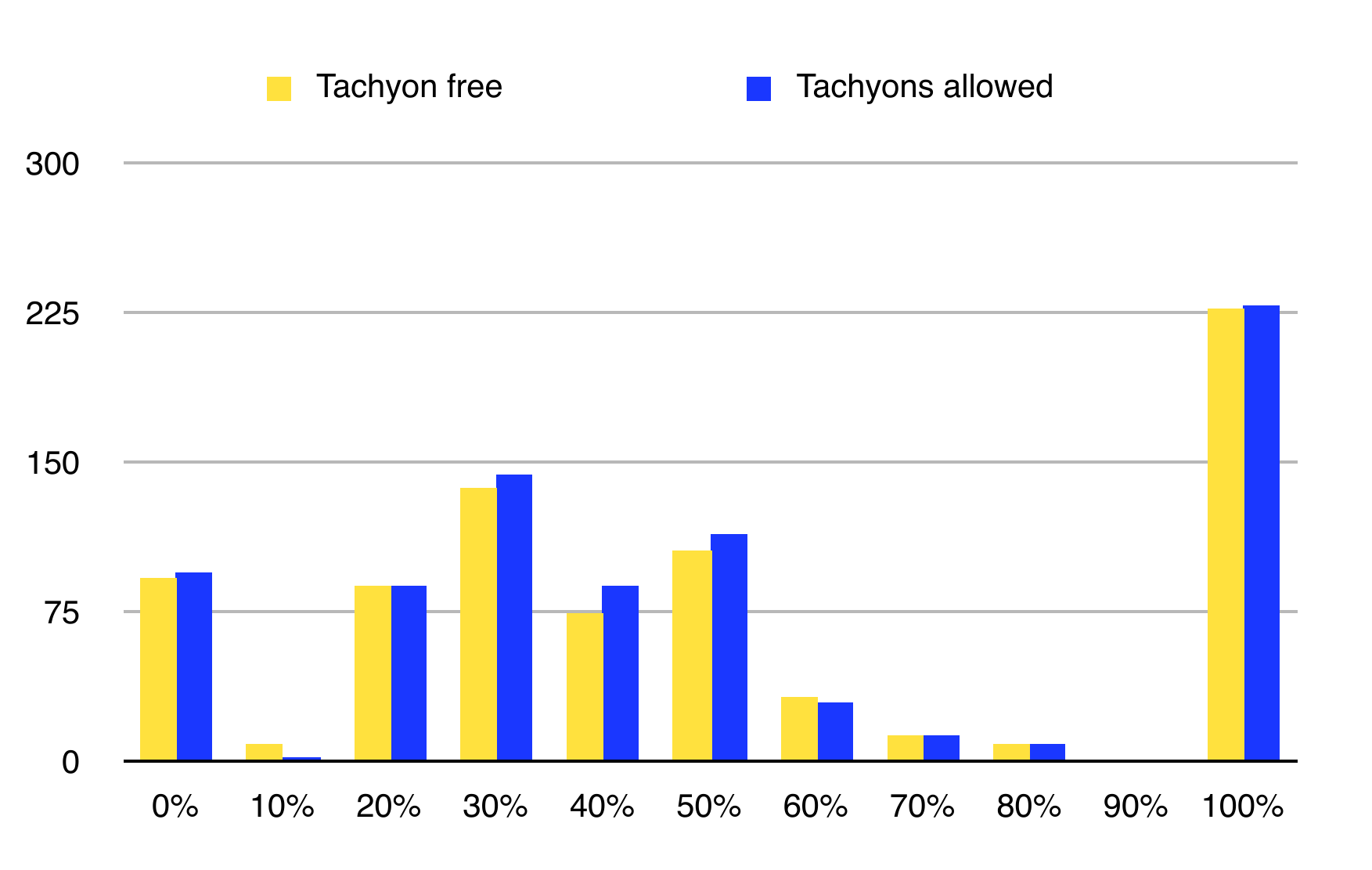}
\label{fig2}
\caption{\em Number of tadpole-free spectra with a given percentage of supersymmetric multiplets. The last bin
consists exclusively of supersymmetric spectra. This plot is for tachyon-removing Klein bottles.}
\end{center}
\end{figure}

If a spectrum is not supersymmetric, is it skewed towards bosons or towards fermions? This is shown in
fig. 3, only for the case of tachyon-free automorphism bulk invariants. What is plotted is the number of bosonic multiplets minus the
number of fermionic multiplet (after subtracting gauginos). The large peak at zero contains  the accidentally supersymmetric
solutions, but of course there are many more where the difference is zero. It appears that there is a very slight preference for a 
surplus of bosons, and that the requirement of absence of tachyons has little effect. The plot for tachyon-free
Klein bottles is similar.

\begin{figure}
\begin{center}
\includegraphics[width=13cm]{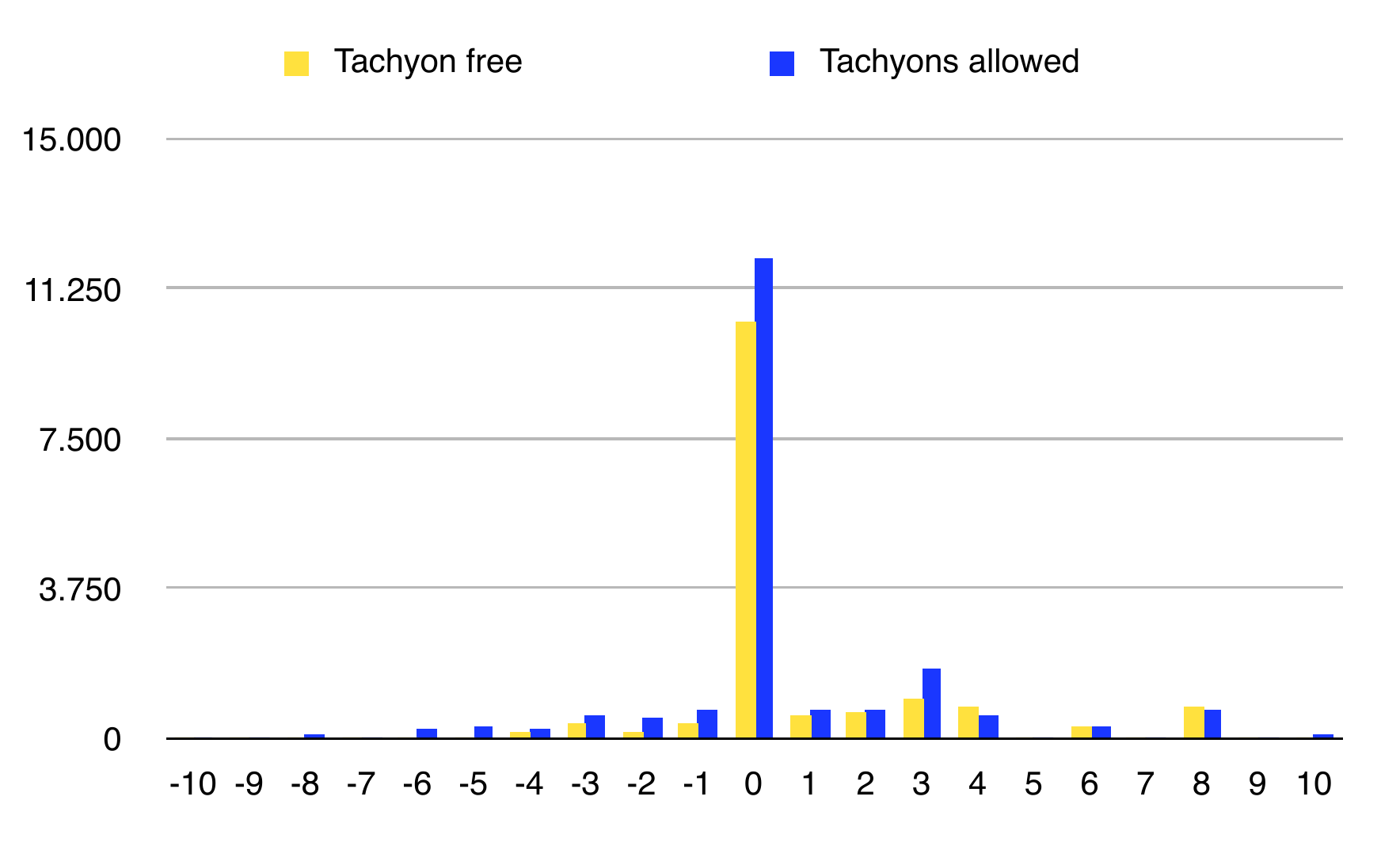}
\label{fig3}
\caption{\em Distribution of boson-fermion surplus for tadpole-free spectra (automorphism case only).}
\end{center}
\end{figure}

In all three cases we did find chiral solutions. In the bulk supersymmetry case, about  $2\%$ of all the fermionic multiplets
appearing in the complete set of solutions is chiral. In the automorphism case this ratio was about $4\%$, in the Klein bottle
case about $1\%$. In figure 4  we plot the  distribution of the chiral multiplets according to net chirality. This plot shows
the same characteristic observed in \cite{Dijkstra:2004cc} and \cite{Gmeiner:2005vz} for the number of 
chiral families: there is a clear dip of a few
orders of magnitude precisely at the
number 3, apparently caused by a combination of two effects: an exponential fall-off with increasing numbers, and
a substantial reduction of odd versus even chiral multiplicities. The cases with chirality 3 are barely visible
in the plot, but they do exist, and there are 43,44, and 0 multiplets respectively.

\begin{figure}
\begin{center}
\includegraphics[width=13cm]{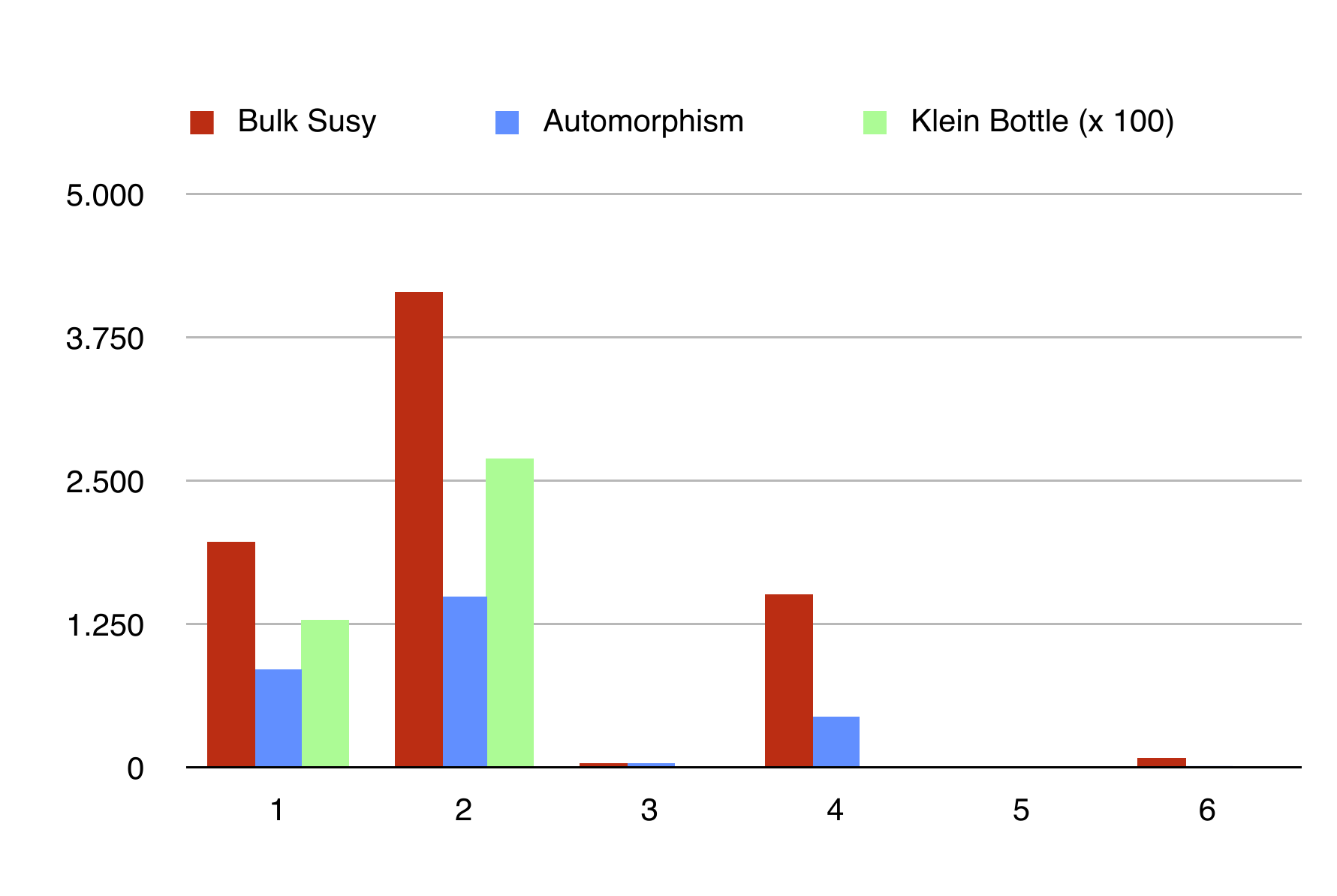}
\label{fig4}
\caption{\em Distribution of chiralities in fermionic multiplets of tachyon and tadpole-free solutions. Zero chirality has been omitted.}
\end{center}
\end{figure}

With regard to our main goal, namely finding spectra that are explicitly non-supersym-metric, tadpole and tachyon-free and
contain the standard model, our conclusion is unfortunately negative: we did not find any such example. However, the
existence of chiral spectra with all these features, except  the last one makes it clear that with enough statistics such examples
must emerge. Of the order of $10^5$ supersymmetric spectra were found in \cite{Dijkstra:2004cc}. Requiring absence of tachyons
in the non-supersymmetric case comes at a price of two  to three orders of magnitude in statistics; requiring full tadpole
cancellation costs a similar factor. So it appears that we must be close to finding just one example. Obviously our chances
would have been quite a bit better by  aiming a bit lower, and searching instead for 1,2 or 4 family models, which are far more
numerous.

The  non-supersymmetric, tadpole and tachyon-free spectra that we do  find have a rather remarkable  tendency to
be partly supersymmetric. We suspect that this originates from the underlying $N=2$ world-sheet supersymmetry. Even
though the bulk theory is not space-time  supersymmetric, it is still possible for the would-be supersymmetric partners of the
characters to pair  up as much as possible, and make the task  of cancelling tadpoles and tachyons more easy. If this is true one
would expect a radically different result if $N=1$ building blocks were used. This should be possible, and we hope to
return to this in the future.


\noindent
{\bf Acknowledgements:}
\vskip .2in
\noindent
We thank Elias Kiritsis for useful discussions. This work has been partially 
supported by funding of the Spanish Ministerio de Ciencia e Innovaci\'on, Research Project
FPA2005-05046, and by the Project CONSOLIDER-INGENIO 2010, Programme CPAN
(CSD2007-00042). The work of A.N.S. has been performed as part of the programs
FP 52  and FP 57 of Dutch Foundation for Fundamental Research of Matter (FOM).

\vskip .5in

\bibliography{REFS}
\bibliographystyle{lennaert}

\end{document}